# Yuen による量子鍵配送の安全性に対する批判とそれ以降

岩越　文尚*

あらまし　量子鍵配送(QKD)は秘密鍵を配送する際に証明可能な安全性を担保する技術として 1984 年から注目を浴びてきた．2005 年には QKD の安全性は，実際に配送される量子状態と配送されるべき理想的な量子状態との間のトレース距離により評価されるべきであるとされてきた．このとき，そのトレース距離には量子鍵配送に失敗する最大確率という解釈が与えられたが，2009 年には H. P. Yuen により前記トレース距離にはそのような意味はないとする批判がなされ，以後，Yuen 自身と O. Hirota, K. Kato, T. Iwakoshi らの批判を経て，2015 年には Iwakoshi により Yuen の批判についての詳細な理由が解説された．その後，Yuen は 2016 年にこれまでの彼の批判をまとめた論文を提出し，量子鍵配送理論が証明可能な安全性を保証するために何が不足しているかについて指摘した．今回はその Yuen の論文の内容とそれ以降のトピックについて解説する．

## 1　はじめに

離れた 2 者，Alice と Bob が秘密のメッセージを交換するため盗聴者 Eve の存在下でいかに暗号鍵を共有するかという課題は，暗号の歴史とともに続いてきた．現代では公開鍵暗号により本課題は解決されているが，公開鍵暗号の安全性は，解読アルゴリズムが向上すれば必然的に脅かされる．

それに対し，情報理論的に証明可能な安全性を保証する手法として 1984 年に登場した[1]量子鍵配送 (Quantum Key Distribution) が注目されている．QKD は，物理法則以外の制約が一切ない盗聴者 Eve がいても配送された鍵が安全であることを証明できるとし，その鍵を One-Time Pad (OTP) に使用することで完全秘匿性を目指す．

OTP の安全性を保証するためには，鍵系列の出現確率が一様独立に任意に近くなければならない．これを保証するのがトレース距離評価量である[2, 3]．この評価量は，配送したい理想的な量子状態と，実際に配送される量子状態の間のトレース距離を小さな量で上から抑えることにより，配送される量子状態を理想的な量子状態に任意に近づけることができ，かつトレース距離は配送される量子状態が理想的な量子状態にならない確率の上限を与えると解釈されてきた[2-6].

ところが 2009 年，H. P. Yuen が「トレース距離はそのような操作的意味を持たない」と指摘した[7]. 以後, Yuen [8-10]に続いて O. Hirota [11, 12], K. Kato [13], T. Iwakoshi [14-18]が批判に続き，最終的には Yuen が導いた結果の意味を Iwakoshi が 2015 年に詳細に解説し[19]，Yuen の批判を真剣に取り上げてもらえるよう呼びかけた．Yuen 自身も 2016 年にそれまでの批判を論文にまとめ，QKD が証明可能な安全性を達成するにはトレース距離の意味だけでなく様々な問題があることも指摘した[20]．このとき Yuen は，量子論の知識がなくても理解できるようにと古典確率論を用いて説明している．

Yuen は KCQ (Keyed Communication on Quantum noise) [21] (Y-00 [22]あるいは量子雑音ストリーム暗号などとも呼ばれる)プロトコルを発案したが，自身のプロトコルを普及させるために QKD を批判しているのではない．QKD は実社会での普及が期待されている技術だが，もし安全性に問題があれば普及前に指摘しなければならないからである．例えばもし，QKD 装置をハッキングする 2010 年の攻撃手法[23]などが，QKD 普及後に発見されていたら，QKD インフラを全て物理的に再構築しなければならなかったであろう．また QKD の研究は，他の物理暗号の安全性の基礎研究にもつながることも理由にある．

本稿では Yuen の主張を整理して説明し，さらにその後の展開についていくつかの知見を紹介する．

## 2　One-Time Pad における完全秘匿性

QKD の目的は，C. E. Shannon により情報理論的安全性

*玉川大学 量子情報科学研究所，東京都 町田市 玉川学園 6-1-1,
t.iwakoshi@lab.tamagawa.ac.jp

が証明されたOTPを実現することにある. 本節ではまずOTPの安全性の定義を述べる[24, 25].

$\boldsymbol{X}$を平文，$\boldsymbol{K}$を共有鍵，$\boldsymbol{C}$を暗号文とする．正規の送信者Aliceは暗号文を$\boldsymbol{C} = \boldsymbol{X} \oplus \boldsymbol{K}$により生成し，正規の受信者Bobは暗号文を$\boldsymbol{X} = \boldsymbol{C} \oplus \boldsymbol{K}$により復号する．このとき，$\boldsymbol{K}$の生起確率が一様独立ならば,

$$\begin{gathered}\Pr(\boldsymbol{X},\boldsymbol{C}) = \Pr(\boldsymbol{X} \mid \boldsymbol{C})\Pr(\boldsymbol{C}) \\ \Pr(\boldsymbol{X},\boldsymbol{C}) = \Pr(\boldsymbol{X})\Pr(\boldsymbol{C}) \\ \therefore \Pr(\boldsymbol{X} \mid \boldsymbol{C}) = \Pr(\boldsymbol{X})\end{gathered} \tag{1}$$

このことは，盗聴者Eveが暗号文$\boldsymbol{C}$を見ても，平文に関するヒントは一切得られないことを意味している．これが情報理論的に安全な暗号における，暗号文単独攻撃に対する平文の完全秘匿性である.

# 3 QKDの安全性証明の定義

QKDにおける$\varepsilon$-securityの定義は次で与えられる[2-5].

$$\tfrac{1}{2}\mathrm{tr}\left|\rho_{\mathrm{ABE}} - \tau_{\mathrm{AB}} \otimes \tau_{\mathrm{E}}\right| \le \varepsilon \tag{2}$$

ここで$\tau_{\mathrm{AB}} \otimes \tau_{\mathrm{E}}$は, 送るべき望ましい量子状態であり, $\rho_{\mathrm{ABE}}$は実際に配送される量子状態である．これらは，AliceとBobが鍵$\boldsymbol{k}_{\mathrm{A}}$, $\boldsymbol{k}_{\mathrm{B}}$を得るとして，次の量子状態で与えられる.

$$\rho_{\mathrm{ABE}} := \sum_{\boldsymbol{k}_{\mathrm{A}},\boldsymbol{k}_{\mathrm{B}}} \Pr(\boldsymbol{k}_{\mathrm{A}},\boldsymbol{k}_{\mathrm{B}}) \left|\boldsymbol{k}_{\mathrm{A}},\boldsymbol{k}_{\mathrm{B}}\right\rangle\left\langle \boldsymbol{k}_{\mathrm{A}},\boldsymbol{k}_{\mathrm{B}}\right| \otimes \rho_{\mathrm{E}}(\boldsymbol{k}_{\mathrm{A}},\boldsymbol{k}_{\mathrm{B}}) \tag{3}$$

$$\tau_{\mathrm{AB}} \otimes \tau_{\mathrm{E}} := \sum_{\boldsymbol{k}} 2^{-|K|} \left|\boldsymbol{k},\boldsymbol{k}\right\rangle\left\langle \boldsymbol{k},\boldsymbol{k}\right| \otimes \tau_{\mathrm{E}} \tag{4}$$

もし$\varepsilon = 0$ならば必然的に$\rho_{\mathrm{ABE}} = \tau_{\mathrm{AB}} \otimes \tau_{\mathrm{E}}$であるが, これはEveが量子状態$\tau_{\mathrm{E}}$を保持していてもAliceとBobの鍵$\boldsymbol{k}$に対するヒントは何も得られないことを意味する．問題は$\varepsilon > 0$の場合である．文献[3-5]の記述を見てみよう.

> "$\varepsilon$ security has an intuitive interpretation: with probability at least $1 - \varepsilon$, the key $S$ can be considered identical to a perfectly secure key $U$, i.e., $U$ is uniformly distributed and independent of the adversary's information. In other words, Definition 1 guarantees that the key $S$ is perfectly secure except with probability $\varepsilon$." [3]

> "In this definition, the parameter $\varepsilon$ has a clear interpretation as the maximum failure probability of the process of key extraction." [4]

> "The above definition of security (Definition 2) has the intuitive interpretation that except with probability $\varepsilon$, the key pair ($S_{\mathrm{A}}$, $S_{\mathrm{B}}$) behaves as a perfect key, as described by (41)." [5]

つまり，攻撃者がどのような情報を入手しようが，鍵の生成分布が一様独立分布$U$になる確率が$1 - \varepsilon$であるとし，そうならない確率$\varepsilon$をもってQKDの失敗確率だと解釈しているのである.

# 4 YuenによるQKDの安全性証明への批判

## 4.1 安全性評価量への批判

一方でYuenは, Eveが盗聴に成功する平均確率$\Pr(\boldsymbol{K}|\boldsymbol{E})$は次のとおりであると指摘した[8, 20].

$$\Pr(\boldsymbol{K} \mid \boldsymbol{E}) \le 2^{-|K|} + \tfrac{1}{2}\mathrm{tr}\left|\rho_{\mathrm{ABE}} - \tau_{\mathrm{AB}} \otimes \tau_{\mathrm{E}}\right| \le 2^{-|K|} + \varepsilon \tag{5}$$

つまり「最大失敗確率」であるトレース距離よりも，Eveが盗聴に成功する確率の方が大きいのである．つまり，「トレース距離はQKDの最大失敗確率である」というステートメント自体が誤りであり，実際[2-5]にはトレース距離こそ最大失敗確率であるという妥当な証明はない.「証明」は2014年に[6]で提出されたが，その証明には送信した量子状態と，送信されてはいないが送信したい理想的な量子状態との間の相関を仮定しており，Katoは「物理的な状況を無視した数学的な操作」と評した[13].

一方で、偶然にもYuenと同じ結果が[6]のAppendixには偶然にもC. PortmannとR. Rennerによって記載されている．それは次のとおりである．トレース距離は一般に(6)の不等号を満たす.

$$\mathrm{tr}\left[\Gamma\left(\varsigma_{\mathrm{ABE}} - \tau_{\mathrm{AB}} \otimes \sigma_{\mathrm{E}}\right)\right] \le \tfrac{1}{2}\mathrm{tr}\left|\varsigma_{\mathrm{ABE}} - \tau_{\mathrm{AB}} \otimes \sigma_{\mathrm{E}}\right| \tag{6}$$

ここで次の測定オペレータを導入する.

$$\Gamma = \sum_{\boldsymbol{k}_{\mathrm{A}}} \left|\boldsymbol{k}_{\mathrm{A}},\boldsymbol{k}_{\mathrm{A}}\right\rangle\left\langle \boldsymbol{k}_{\mathrm{A}},\boldsymbol{k}_{\mathrm{A}}\right| \otimes M_{\boldsymbol{k}_{\mathrm{A}}} \tag{7}$$

したがって,

$$\begin{aligned}&\mathrm{tr}\left[\Gamma\left(\varsigma_{\mathrm{ABE}} - \tau_{\mathrm{AB}} \otimes \sigma_{\mathrm{E}}\right)\right] \\ &= \sum_{\boldsymbol{k}_{\mathrm{A}}}\left[p(\boldsymbol{k}_{\mathrm{A}},\boldsymbol{k}_{\mathrm{A}})\Pr\left(\boldsymbol{k}_{\mathrm{A}} \mid \boldsymbol{E}\right)\right] - 2^{-|K|} \\ &= \Pr(\boldsymbol{K} \mid \boldsymbol{E}) - 2^{-|K|}\end{aligned} \tag{8}$$

$2^{-|K|}$を移項すれば(5)が得られる．この手続は、量子情報の入門的な書籍[26]を見れば記載されている。

ここで気をつけて頂きたいのは，(5)の導出は，特定のプロトコルや実装，証明手順に依存しないことである.

トレース距離によりQKDの安全性を評価するかぎり，全ての安全性理論に当てはまる．「量子暗号に30年ぶりの新原理 –「読まれたら気づく」から「読めない」手法へ」とプレスリリース[27]されたRound Robin DPS QKD [28, 29]にさえもこの論法は成立する．

また，Yuenは定量的な安全性にもコメントしている．現時点ではQKDにおいて配送される鍵長は標準的には $|K| = 10^6$ bitsであるとされている．これに対し，現時点で最も小さな $\varepsilon$ [29]を参照してもたかだか

$$\Pr(\boldsymbol{K} \mid \boldsymbol{E}) \sim \varepsilon = 2^{-50} \gg \Pr(\boldsymbol{K}) = 2^{-|K|} = 2^{-1,000,000} \quad (9)$$

つまり，OTPの完全秘匿性である $\Pr(\boldsymbol{K}|\boldsymbol{E}) = \Pr(\boldsymbol{K})$ を全く満たさず，OTPが必要とする鍵の生起確率の一様独立性を担保できないことである．

もちろん $\varepsilon$ が十分に小さければ $\Pr(\boldsymbol{K}|\boldsymbol{E})$ が十分に小さくなり，QKDで配送された鍵の安全性も担保される．だが，$\Pr(\boldsymbol{K}|\boldsymbol{E}) \sim \varepsilon = 2^{-50}$ 程度では，2008年に日本で起きた自動車1台あたりの年間死亡確率とQKDが盗聴される潜在的確率 $\Pr(\boldsymbol{K}|\boldsymbol{E})$ はほぼ同じである[19]．一方で，仮にAESを効率的に解読できるアルゴリズムがなかったとすると，256 bitの鍵長のAESを当てずっぽうで解読に成功する確率は $2^{-256}$ であり，YuenやHirotaが「QKDはAESより弱い」と何度か説明してきたことの意味がわかる．だがQKDの理論によれば，$\varepsilon$ は任意に小さくできるとのことなので，今後の研究に注視したい．ただし，$\varepsilon$ を小さくすれば鍵配送速度が低下し，場合によっては $\varepsilon$ をある量以下にできないというトレードオフがある[30]．

## 4.2 QKD後のOne-Time Padへの既知平文攻撃

鍵の生成確率分布が一様独立でないならば，鍵の一部を知ることにより，他の部分の推測に成功する確率も高まる．特にOTPは，暗号文を平文と鍵系列との排他的論理和で生成するので，もし平文を一部推測ができればその部分の鍵系列はすぐに明らかになる．

既存のQKDの安全性の理論では，3節で述べたように，高確率で一様独立分布の鍵系列が生成されるため，既知平文攻撃は不可能であるとされていた．そもそも，[3]がトレース距離評価量を導入した目的は，既知平文攻撃を無効化することであったし，[6]も既知平文攻撃は不可能だとの証明を記載している．

しかしYuenは，(5)と同じ論法で既知平文攻撃が可能であることを示し，その成功確率を算出した[20]．量子論的には，次のようにして導くことができる．まず(3)において $\boldsymbol{k}_{\mathrm{A}}$ をEveにとって既知の部分 $\boldsymbol{k}_{\mathrm{A1}}$ と未知の部分 $\boldsymbol{k}_{\mathrm{A2}}$ とに分割し，既知の部分 $\boldsymbol{k}_{\mathrm{A1}}$ を廃棄すると，

$$\rho_{\mathrm{ABE}} := \sum_{\boldsymbol{k}_{\mathrm{A2}},\boldsymbol{k}_{\mathrm{B}}} \Pr(\boldsymbol{k}_{\mathrm{A2}},\boldsymbol{k}_{\mathrm{B}}) \left|\boldsymbol{k}_{\mathrm{A2}},\boldsymbol{k}_{\mathrm{B}}\right\rangle\left\langle \boldsymbol{k}_{\mathrm{A2}},\boldsymbol{k}_{\mathrm{B}}\right| \otimes \rho_{\mathrm{E}}(\boldsymbol{k}_{\mathrm{A2}},\boldsymbol{k}_{\mathrm{B}}) \quad (10)$$

$$\Gamma = \sum_{\boldsymbol{k}_{\mathrm{A2}}} \left|\boldsymbol{k}_{\mathrm{A2}},\boldsymbol{k}_{\mathrm{A2}}\right\rangle\left\langle \boldsymbol{k}_{\mathrm{A2}},\boldsymbol{k}_{\mathrm{A2}}\right| \otimes M_{\boldsymbol{k}_{\mathrm{A2}}} \quad (11)$$

残った系について鍵の推定に成功する確率を計算すると(9)と同様に

$$\Pr(\boldsymbol{K}_{\mathrm{A2}} \mid \boldsymbol{E}) \le 2^{-|K_{\mathrm{A2}}|} + \varepsilon \quad (12)$$

以上から既知平文攻撃による鍵の推定は無意味ではないことが示された．しかし本執筆者の意見を述べれば，$\varepsilon$ の大きさには影響がないため，(9)に示したように $\varepsilon$ が無視できないほど大きい場合には，鍵を推定することについての既知平文攻撃は暗号文単独攻撃と大差はないとも言える．逆に $\varepsilon$ が $2^{-|K|}$ 程度であれば，既知平文攻撃は意味を持ってくる．

## 4.3 盗聴者にとってのビット誤り率の重要性

Yuenは次のような疑問を提起している．例えばEveが正しい鍵を推定できなかったとしても，それとは1 bit違う鍵で暗号文を復号したらEveにとって復号した平文は全く読めないものだろうか？2 bitならばどうか？3 bitならば？ Yuenは，Eveにとっての鍵ビットの推定誤り率が平文の秘匿性に重要であると考え，これをBit-Error-Rate Guaranteeと呼んでいる[9, 20]．当然，OTPにとって理想的な一様独立な生成確率をもつ鍵系列ならばEveにとってのビット誤り率は1/2であり，Eveは平文を全く推定できない．しかし1/2より十分小さければ部分的にでも平文を読める可能性が生じる．例えば，我々が知人から「ありかとう」という文字列を受け取ったとしよう．このとき我々は，「ありがとう」の書き損じだったのだと普通ならば推測する．

QKDの話に戻って，以下は本執筆者の概算であるが，正規ユーザーが共有している鍵系列から $Q_{\mathrm{E}}$ までの誤り率まで含んだ鍵系列までEveが平文を読める範囲に含まれるとすると，Eveが平文を読むことができる確率は，大雑把に見積もって

$$\Pr(K \mid E) 2^{|K| h_2(Q_{\mathrm{E}})} \le 2^{-|K|(1-h_2(Q_{\mathrm{E}}))} + \varepsilon 2^{|K| h_2(Q_{\mathrm{E}})} \quad (13)$$

ここで $h_2(Q_{\mathrm{E}})$ は $Q_{\mathrm{E}}$ を確率とするShannon Binary Entropyである．このようにEveが平文を読むことができる確率は，鍵の長さに対して指数関数的に増加する．もし，$\varepsilon = 2^{-50}$ と $|K| = 10^6$ bit対して，Eveにとって $Q_{\mathrm{E}} = 2.5\times10^{-6}$ までは復号した平文を読むのに問題はないとすると，(13)の右辺は1となり，確実に盗聴されることになる．実際，このときEveが推定した $10^6$ bitの鍵の中に含まれる誤りビット数はたかだか2, 3 bitである．もちろん(13)の計算は大雑把すぎるであろうし，どの程度の $Q_{\mathrm{E}}$ ならば平文の秘匿性が保たれるのかはYuenもOpen Questionであるとしている．今後の研究が必要である．

### 4.4　鍵配送レート算出法への批判

Yuen は鍵の生成レートの計算方法についても疑問を提起している．QKD の手順はよく知られているが，以下に述べておく[31].

1. 送信者 Alice は送信ビットと送信基底をランダムに選び，対応する量子状態を受信者 Bob に送る.
2. Bob はランダムに測定基底を選び，受信した量子状態から受信ビットを得る.
3. Alice と Bob は使用した基底のみを古典回線で公開し，基底が一致しないビットを破棄，一致したものを保持する.
4. Alice と Bob は保持しているビット列の一部を公開し，量子誤り率 $Q$ を計算する．$Q$ が所定以上ならば，安全な鍵を生成できないので通信を終了する．所定以下であれば誤り訂正を行う.
5. Alice はパリティ検査行列を公開し，さらにこの行列を用いて彼女の鍵のシンドロームを計算する．計算されたシンドロームはあらかじめ共有している鍵を消費して OTP で Bob に伝えられる.
6. Bob も公開されたパリティ検査行列から彼の鍵列のシンドロームを計算し，Alice のものと比較することで誤り訂正を行う.
7. その後さらに Privacy Amplification により鍵を hash するため，これに対応する行列を公開して演算し，最終鍵を手にする.

上記の誤り訂正でシンドロームを隠すために用いる OTP の鍵の消費量は，次の形式で表現される.

$$\mathrm{leak}_{\mathrm{EC}} = \xi k h_2(Q) \tag{15}$$

ここで $\xi$ は誤り訂正コードの効率により決まるとされており，1 から 2 までの値から選ばれ，たいていは習慣的に $\xi = 1.1$ とされることが多い[31, 32].

$\xi = 1$ の場合の (15) は，次のように導出される．まず，コード長 $k$ で情報ビット長 $m$ の線形$(k, m)$符号が $Qk$ bits までの誤りを訂正できるとする．ハミングの不等式より

$$2^m \le 2^k / \sum_{a=0}^{Qk} {}_k C_a \tag{16}$$

さらに

$$\sum_{a=0}^{Qk} {}_k C_a \le 2^{kh_2(Q)} \tag{17}$$

このことから次を満たせばよいことがわかるので

$$\sum_{a=0}^{Qk} {}_k C_a \le 2^{kh_2(Q)} \le 2^{k-m} \tag{18}$$

よって，OTP で隠すべきシンドロームの長さは $k\ h_2(Q)$ bits となる.

しかし Yuen は次のように説明する．$(k, m)$ 符号により誤り訂正すると，訂正前は $2^k$ とおりの可能性があった鍵のパターンが $2^m$ とおりに縮小する．この問題は OTP でシンドロームを隠していても防げない．この問題を解決するには，保持している $k$ bit の鍵にパリティ検査符号を足した $(n, k)$ 符号を考え，OTP で隠すパリティ検査符号の長さを $n\, h_2(Q)$ bit とする必要がある．$n$ の長さは，再びハミングの不等式より

$$\sum_{a=0}^{Qn} {}_n C_a \le 2^{nh_2(Q)} \le 2^{n-k} \tag{19}$$

よって，OTP に使用する鍵の消費量は

$$\mathrm{leak}_{\mathrm{EC}} = kh_2(Q) / (1 - h_2(Q)) \ge kh_2(Q) \tag{20}$$

さらに Yuen は(15)について、習慣的に $\xi = 1.1$ とすることに対し「QKD は証明された安全性を提供するという主張に反する」としている.

ここで，本執筆者が数値計算で得た結果[30]の一部を Fig. 1 に示す．Yuen が提案した式では，$Q$ が大きくなると鍵生成レートに強く制限がかかることになる.

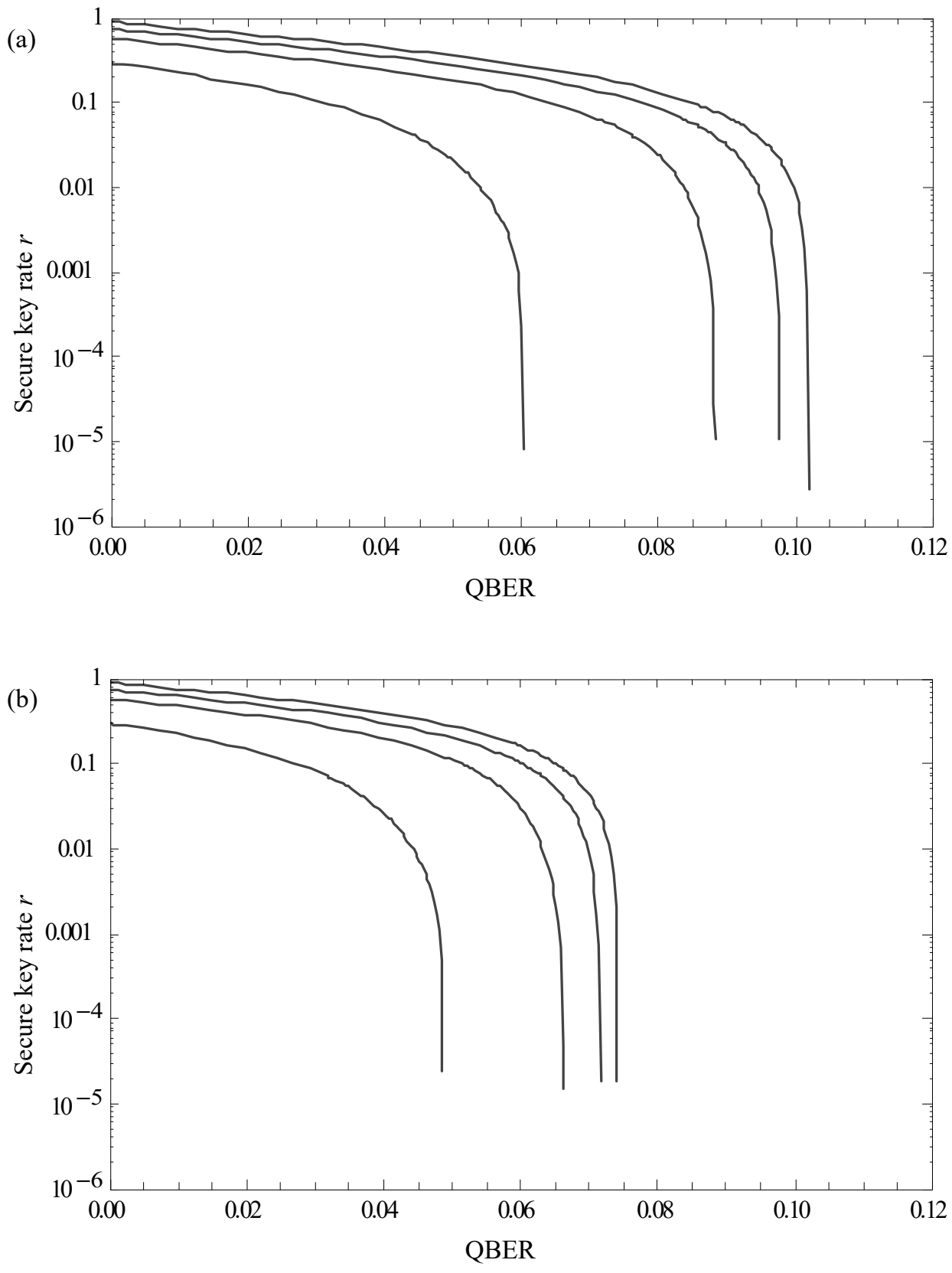

Fig. 1(a). Key rate with $\mathrm{leak}_{\mathrm{EC}}$ in (15), and (b) with $\mathrm{leak}_{\mathrm{EC}}$ in (20).

From the lowest curve, the sifted key length = $10^5$, $10^6$, $10^7$, $10^9$ bits.

### 4.5　Privacy Amplification への批判

Yuen は Privacy Amplification の有用性についても疑問を呈している．Yuen の説明では理解しにくいので，ここで本著者が別の例を述べる．Eve は量子回線を盗聴して量子メモリに情報を蓄える．さらに Alice と Bob が誤り訂正を行ったあと，この時点で彼らが共有した鍵を Eve が推定したとする．Eve が推定した鍵を $\boldsymbol{k}_{\mathrm{E}}$，Alice と Bob が共有した鍵を $\boldsymbol{k}_{\mathrm{R}}$ としよう．Privacy Amplification にどのような hash 関数 $f$ を用いるかを Alice は公開し，両者はそれに従って hashing を行う．Hash 関数が(21)のような $\delta$-Almost Two-Universal であるなら，

$$\Pr_{f\in F}(f(\boldsymbol{k}_{\mathrm{E}})=f(\boldsymbol{k}_{\mathrm{R}}))\le\delta \tag{21}$$

Eve が正しく $\boldsymbol{k}_{\mathrm{E}}=\boldsymbol{k}_{\mathrm{R}}$ を選んだ確率と，Eve が $\boldsymbol{k}_{\mathrm{E}}\ne\boldsymbol{k}_{\mathrm{R}}$ を選んだにもかかわらず hash 後の衝突が起きる可能性を考えなくてはならないので，Eve が Privacy Amplification 後に鍵の推定に成功する確率は,

$$\begin{aligned}P_{\mathrm{guess}}&=\Pr(\boldsymbol{k}_{\mathrm{E}}=\boldsymbol{k}_{\mathrm{R}})+\Pr(\boldsymbol{k}_{\mathrm{E}}\ne\boldsymbol{k}_{\mathrm{R}})\Pr(f(\boldsymbol{k}_{\mathrm{E}})=f(\boldsymbol{k}_{\mathrm{R}}))\\&\le\Pr(\boldsymbol{k}_{\mathrm{E}}=\boldsymbol{k}_{\mathrm{R}})+\left(1-\Pr(\boldsymbol{k}_{\mathrm{E}}=\boldsymbol{k}_{\mathrm{R}})\right)\delta\\&=(1-\delta)\Pr(\boldsymbol{k}_{\mathrm{E}}=\boldsymbol{k}_{\mathrm{R}})+\delta\end{aligned} \tag{21}$$

この右辺は hash 関数を適用する前の確率よりも大きい．すると次のような疑問が起こる．Privacy Amplification は本当に配送された鍵の安全性を高めているのだろうか？ Hashing 前の鍵の長さを$|K_{\mathrm{R}}|$，hashing 後の鍵の長さを$|K|$とし，$|K_{\mathrm{R}}|>|K|$とする．Eve が仮にまったく盗聴せずに，当てずっぽうで鍵を推測したとしても，hashing 後の鍵のほうが当てやすいのは自明である．Yuen は，Privacy Amplification で安全性が高まるとの結論が導かれた理由を，Leftover Hash Lemma において hash 関数の family $F$ からランダムに選ばれた $f$ で結果を平均化していることを挙げている．しかし実際には Alice と Bob は $f$ を公開するので，Eve はどの $f$ が使われたのかを知っている．この具体的な $f$ に対して安全性評価を行う必要がある．

さらにやっかいな問題は，前述の誤り訂正と関連する．誤り訂正前の鍵を codeword だとして誤り訂正を行えば，鍵列のビットには特定のパターンが生じる．なぜならば $k$ bit の鍵を$(k, m)$符号としているので、実際には鍵は $2^k$ 通りのランダムな系列ではなく $2^m$ 通りに限られた系列だからである．この状態で Privacy Amplification 後の鍵の安全性を評価するのは簡単ではない．やはり前述のようにパリティチェック符号を鍵に足して鍵列のビット間の独立性をなるべく保持したほうがよさそうであり前述の誤り訂正の問題は真剣に考えるべき課題であると言える．

### 4.6　通信路の認証鍵の安全性

QKD の研究者の間では共有された前提なのであまり説明されておらず，誤解されている方々を散見するのでここで説明しておく．Alice と Bob は互いに通信相手が自分であることを証明するための認証鍵が必要である[31, 33, 34]. これがなければ Eve は Alice と Bob になりすまし，量子・古典とも全ての信号を中継することで盗聴ばかりでなく平文の改ざんまで行える中間者攻撃を実行できる．この意味で QKD は RSA 暗号に代わる技術には完全にはなりえない．RSA 暗号の公開鍵はもちろん Eve も知っているが QKD の認証鍵は Eve に知られてはならないからである．従って QKD を始める前には認証鍵が秘密裏に共有されていなければならない．つまり QKD は，どちらかと言えば AES と同じ共通鍵暗号なのである．

Yuen はこの認証鍵の安全性についても警鐘を鳴らしている．最初に共有する鍵は一様独立な確率分布をもつと仮定してもよいであろうが，QKD により配布された鍵で認証鍵を更新するとどのようになるであろうか．先に述べたように，Eve は鍵を $\varepsilon$ 程度の確率で推測できる．Yuen は，認証は暗号化よりも重要な課題であり，さらに小さな $\varepsilon$ で安全性を評価できなければならない，と説明している．さらに $\varepsilon$ が十分に小さかったとしても，完全に一様独立な確率分布をもった最初の認証鍵に比べ，Eve が確率 $\varepsilon$ で推測できる鍵の一部を認証鍵として使う次の QKD のラウンドの安全性への影響を評価するのは難しい．なお、Eve が確率 $\varepsilon$ で推測できる鍵は、前述の誤り訂正の際に使われるシンドロームを隠すための OTP にも使われるので、この安全性への影響も評価する必要がある。

## 5　量子暗号の近年の動向

アメリカ国防高等研究計画局 (DARPA) は，2012 年に量子暗号に次の要求を課した[35].

・通信速度: 1-10 Gbps
・通信距離: 1,000-10,000 km

これは QKD ではかなり困難な課題である．

また，イギリス政府の情報機関である CESG の白書では，QKD の開発からの撤退が提案されている[36]．その理由は下記のとおりである．

1. QKD の研究は暗号化が中心であり，認証や署名などの他のセキュリティの研究があまりなされていない．
2. QKD は Point-To-Point の通信であり，通信距離にも制限がある．さらにネットワークを構築するには,

古典物理学で保護された中継点を設ける必要があり，量子力学で保護された無条件安全とはならない．
3. QKD 装置への攻撃手法[23, etc.]も研究されているが，こうした研究で完全な安全性が保証されるわけではない．未知のループホールが存在する可能性はある．
4. QKD に対する代替案が出ており，例えば耐量子計算暗号などはソフトウェアで実装できるが，QKD 装置のアップデートはハードウェア交換が必要である．
5. 従って現在のところはソフトウェアで実装できる耐量子計算暗号の開発を優先することを推奨する．

一方，QKD 以外の量子暗号もいくつか発表されている．まず Yuen 自身が考案した KCQ [21] (Y-00 [22]や量子雑音ストリーム暗号などとも呼ばれ，玉川大学で改良されたものは量子エニグマ暗号と名付けられた[37])，S. Lloyd の量子エニグマママシン[38]，J. H. Shapiro が提案した量子イルミネーションを応用した暗号[39]である．いずれも単一光子のような微弱な光ではなく，現在の光通信と同程度の強度の光信号の巨視的量子性を用いた暗号である．

話を KCQ に限定すると，よくある誤解で KCQ は初期鍵を用いて平文を直接暗号化する方式で，鍵を配送することが目的ではないと言われるが，単に平文の代わりに新しい鍵を送信するだけで，QKD と同様に鍵配送が可能である．以前から「QKD で配送した鍵を KCQ で用いる」という組み合わせが提案されてきたが，QKD の安全性がさらに向上されない限りはあまり意味のない組み合わせであろう．

もちろん，量子力学により古典通信では不可能だった秘匿通信を QKD という形で実現できることを示唆した先人たちの基礎科学的な功績は讃えられるべきである．さらに本稿で述べた通り，QKD は物理暗号に要求される安全性とは何かという重大な課題も浮き彫りにしてくれている．しかし実用性という観点から鑑みて，QKD 以外の方策も研究される価値は十分にあると思われる．そのときでも，QKD で研究された知見が生かされる可能性は十分にあると執筆者は考える．

## 6 まとめ

量子鍵配送 (QKD) は秘密鍵を配送する際に証明可能な安全性を保証する技術として 1984 年から注目を浴びてきた．しかし 2009 年からは H. P. Yuen が安全性証明に疑問を提起し，以後，Yuen 自身と O. Hirota，K. Kato, T. Iwakoshi らによる解説が続けられた．その後，Yuen は 2016 年にこれまでの彼の批判をまとめた論文を提出し，QKD が証明可能な安全性を保証するには何が不足しているかについて指摘した．本報告では Yuen が指摘した QKD の問題点を解説し，かつ近年の QKD 以外の暗号化方式を模索する動向について述べた．QKD それ自体は，量子力学を用いた秘匿通信が可能だということを示した重要な意義のある研究である．しかし Yuen により QKD の安全性理論の問題点も明らかになり，QKD 以外の通信方式を研究するという選択肢も提案されている．その際であっても，QKD の研究で得られた知見が役に立つ可能性は十分にあると執筆者は考える．

## 参考文献

# Yuen's Criticisms on Security of Quantum Key Distribution and Onward

Takehisa Iwakoshi*

**Abstract** Quantum Key Distribution (QKD) has been attracting researchers that it would provide provable security to distribute secret keys since its birth in 1984. Since 2005, the trace distance between an ideal quantum state and an actually distributed state has been employed to evaluate its security level, and the trace distance was given an interpretation that it would be a maximum failure probability in distributing perfectly secure keys. However, in 2009, H. P. Yuen criticized that the trace distance would not have such an interpretation. Since then, O. Hirota, K. Kato, and T. Iwakoshi have been warning to make people pay attention to Yuen's criticisms. In 2015, T. Iwakoshi precisely explained why Yuen has been correct. In 2016, Yuen himself published a paper to explain the potentially unsolved problems in QKD. This study precisely explains the most important problems given in Yuen's paper, and gives recent topics around QKD and other quantum cryptographic protocols.

## 7 Introduction

There has been the most important problem in cryptology that two separated parties, say, Alice and Bob, have to share the secret key to exchange private messages under the presence of an eavesdropper, Eve, from the beginning of the birth of Cryptography. Nowadays, this problem was solved by Public-Key Encryption protocols. However, such Public-Key Encryptions will be threatened when better decryption algorithms are developed.

On the other hand, Quantum Key Distribution (QKD) was proposed in 1984 [1] claiming it would enable to provide provably secret keys under the presence of Eve with unlimited power other than the laws of physics. Then, the secret keys would realize One-Time Pad (OTP), which was proven to be perfectly secure.

To guarantee the security of OTP, the probability distribution of a key string has to be arbitrarily close to an Independent and Identically Distribution (IID). In case of QKD, this would be guaranteed by the trace distance between an ideal quantum state and an actually distributed state with an upper-bound of arbitrarily small number [2, 3]. This criterion has been perceived that it allows making the two quantum states arbitrarily close and the trace distance itself would be a maximum failure probability in distributing perfectly secure keys for OTP [2-6].

However, in 2009, H. P. Yuen pointed out that the trace distance would not have such a maximum failure probability interpretation [7]. Since Yuen continued his criticisms [8-10], O. Hirota [11, 12], K. Kato [13], T. Iwakoshi [14-18] have been warning that Yuen would be correct, and finally in 2015, Iwakoshi explained precisely why Yuen was correct and attentions must be paid to Yuen's claims seriously [19]. Yuen himself published his invited paper from IEEE in 2016 to summarize his criticisms, and explained there are many unsolved problems to say QKD is provably secure, not only the interpretation of the trace distance [20]. In the literature, Yuen explained using classical probabilities so that conventional cryptologists can understand without knowing the details of Quantum Mechanics.

Note that, his criticisms have been done not because Yuen aims to standardize his own quantum encryption protocol, which was named Keyed Communication in Quantum-noise (KCQ) at first and now which is called Y-00 or Quantum noise stream cipher. QKD is expected to be the world-widely important network infrastructure. Therefore, all problems have to be checked carefully before QKD is established as physical network infrastructures. Imagine the hacking attack in 2010 [23] was found after the QKD infrastructures were established; we would have required to replace all QKD infrastructures physically. Other than that, it is useful to study what is required for physical encryption systems through the studies on QKD.

*Quantum ICT Research Institute, Tamagawa Univ., 6-1-1 Tamagawa-Gakuen, Machida, Tokyo 194-8610, Japan. t.iwakoshi@lab.tamagawa.ac.jp

Therefore, this paper explain precisely why Yuen has been warning, and explain the recent topics related quantum cryptography.

## 8 Perfect Secrecy of One-Time Pad

The goal of QKD is to provide key strings for OTP which was proven to be perfectly secure by C. E. Shannon. This section shows what was the definition of perfect secrecy along Shannon's contexture [24, 25].

Let $\boldsymbol{X}$ be a plaintext string, let $\boldsymbol{K}$ be a shared secret key string, and let $\boldsymbol{C}$ be a ciphertext string. The legitimate transmitter, Alice, generates $\boldsymbol{C}$ by $\boldsymbol{C} = \boldsymbol{X} \oplus \boldsymbol{K}$, then the legitimate receiver Bob decode $\boldsymbol{C}$ by $\boldsymbol{X} = \boldsymbol{C} \oplus \boldsymbol{K}$. If the probability distribution of $\boldsymbol{K}$ is IID, then

$$\begin{gathered} \Pr(\boldsymbol{X},\boldsymbol{C}) = \Pr(\boldsymbol{X} \mid \boldsymbol{C})\Pr(\boldsymbol{C}) \\ \Pr(\boldsymbol{X},\boldsymbol{C}) = \Pr(\boldsymbol{X})\Pr(\boldsymbol{C}) \\ \therefore \Pr(\boldsymbol{X} \mid \boldsymbol{C}) = \Pr(\boldsymbol{X}) \end{gathered} \tag{1}$$

This means Eve never obtains any knowledge on the plaintext $\boldsymbol{X}$ no matter how much she obtains the ciphertext $\boldsymbol{C}$. This is the definition of the perfect secrecy against Ciphertext-Only Attack on information theoretic secure cryptographies.

## 9 Security definitions in QKD

QKD employs a concept "ε-security" defined as [2-5]

$$\tfrac{1}{2}\mathrm{tr}\left|\rho_{\mathrm{ABE}} - \tau_{\mathrm{AB}} \otimes \tau_{\mathrm{E}}\right| \le \varepsilon \tag{2}$$

Here, $\tau_{\mathrm{AB}} \otimes \tau_{\mathrm{E}}$ is an ideal quantum state to be distributed, $\rho_{\mathrm{ABE}}$ is a quantum state actually distributed. In more details, with $\boldsymbol{k}_{\mathrm{A}}$ and $\boldsymbol{k}_{\mathrm{B}}$ to be Alice's and Bob's key strings respectively, these quantum states are defined as

$$\rho_{\mathrm{ABE}} := \sum\nolimits_{\boldsymbol{k}_{\mathrm{A}},\boldsymbol{k}_{\mathrm{B}}} \Pr(\boldsymbol{k}_{\mathrm{A}},\boldsymbol{k}_{\mathrm{B}}) \left|\boldsymbol{k}_{\mathrm{A}},\boldsymbol{k}_{\mathrm{B}}\right\rangle\left\langle \boldsymbol{k}_{\mathrm{A}},\boldsymbol{k}_{\mathrm{B}}\right| \otimes \rho_{\mathrm{E}}(\boldsymbol{k}_{\mathrm{A}},\boldsymbol{k}_{\mathrm{B}}) \tag{3}$$

$$\tau_{\mathrm{AB}} \otimes \tau_{\mathrm{E}} := \sum\nolimits_{\boldsymbol{k}} 2^{-|K|} \left|\boldsymbol{k},\boldsymbol{k}\right\rangle\left\langle \boldsymbol{k},\boldsymbol{k}\right| \otimes \tau_{\mathrm{E}} \tag{4}$$

If $\varepsilon = 0$, $\rho_{\mathrm{ABE}} = \tau_{\mathrm{AB}} \otimes \tau_{\mathrm{E}}$ is satisfied necessarily, therefore it is meaningless for Eve to have a quantum state $\tau_{\mathrm{E}}$ to guess $\boldsymbol{k}_{\mathrm{A}}$ and $\boldsymbol{k}_{\mathrm{B}}$ because $\tau_{\mathrm{E}}$ is not correlated to their state $\tau_{\mathrm{AB}}$. The problem is in the case of $\varepsilon > 0$. Now we refer descriptions in [3-5].

> "ε security has an intuitive interpretation: with probability at least 1 – ε, the key $S$ can be considered identical to a perfectly secure key $U$, i.e., $U$ is uniformly distributed and independent of the adversary's information. In other words, Definition 1 guarantees that the key $S$ is perfectly secure except with probability ε." [3]

> "In this definition, the parameter ε has a clear interpretation as the maximum failure probability of the process of key extraction." [4]

> "The above definition of security (Definition 2) has the intuitive interpretation that except with probability ε, the key pair ($S_{\mathrm{A}}$, $S_{\mathrm{B}}$) behaves as a perfect key, as described by (41)." [5]

This means, the trace distance is interpreted as the maximum failure probability in distributing the key string with its probability distribution is IID whatever Eve's quantum state is; the failure probability in distributing the perfect key string is ε, and with a probability of 1 – ε, QKD succeeds in distributing perfectly secure key strings.

## 10 Yuen's criticisms on QKD security proofs

### 10.1 Criticisms on security evaluation

Yuen wrote that Eve's average success probability in eavesdropping, $\Pr(\boldsymbol{K}|\boldsymbol{E})$, is [8, 20]

$$\Pr(\boldsymbol{K} \mid \boldsymbol{E}) \le 2^{-|K|} + \tfrac{1}{2}\mathrm{tr}\left|\rho_{\mathrm{ABE}} - \tau_{\mathrm{AB}} \otimes \tau_{\mathrm{E}}\right| \le 2^{-|K|} + \varepsilon \tag{5}$$

The above shows that Eve's success probability can be larger than "the maximum failure probability." This means that the statement "the upper-bound of the trace distance itself is the failure probability" is incorrect. Actually, Refs. [2-5] have no valid proofs that the trace distance itself is the maximum failure probability. The "proof" was given in 2014 by Ref. [6], however, it assumed there had to be a correlation between the distributed quantum state and an ideal quantum state which is not distributed in reality. Kato pointed out that this calculation is mathematically valid but it ignored physical restrictions [13].

The proof of (5) was coincidentally written in Appendix of Ref. [6] by C. Portmann and R. Renner, too. It is as follows. A trace distance satisfies the following inequality (6) in general.

$$\mathrm{tr}\left[\Gamma\left(\varsigma_{\mathrm{ABE}} - \tau_{\mathrm{AB}} \otimes \sigma_{\mathrm{E}}\right)\right] \le \tfrac{1}{2}\mathrm{tr}\left|\varsigma_{\mathrm{ABE}} - \tau_{\mathrm{AB}} \otimes \sigma_{\mathrm{E}}\right| \tag{6}$$

Then, introduce a projection operator as follows.

$$\Gamma = \sum\nolimits_{\boldsymbol{k}_{\mathrm{A}}} \left|\boldsymbol{k}_{\mathrm{A}},\boldsymbol{k}_{\mathrm{A}}\right\rangle\left\langle \boldsymbol{k}_{\mathrm{A}},\boldsymbol{k}_{\mathrm{A}}\right| \otimes M_{\boldsymbol{k}_{\mathrm{A}}} \tag{7}$$

Therefore,

$$\begin{aligned}&\mathrm{tr}\left[\Gamma\left(\varsigma_{\mathrm{ABE}}-\tau_{\mathrm{AB}}\otimes\sigma_{\mathrm{E}}\right)\right]\\&=\sum_{\boldsymbol{k}_{\mathrm{A}}}\left[p(\boldsymbol{k}_{\mathrm{A}},\boldsymbol{k}_{\mathrm{A}})\Pr\left(\boldsymbol{k}_{\mathrm{A}}\mid\boldsymbol{E}\right)\right]-2^{-|K|}\\&=\Pr(\boldsymbol{K}\mid\boldsymbol{E})-2^{-|K|}\end{aligned}\tag{8}$$

By transposing $2^{-|K|}$ in (8), we recover (5). The proof of (6) is written in standard textbooks, for instance, Ref. [26].

Note that the derivation of (5) is independent of the QKD protocols, security proof steps, or implementation of the QKD systems. It is applicable to the all QKD proofs as long as the security can be evaluated by the trace distance; it is applicable even to the protocol press-released as "First breakthrough in quantum cryptography in 30 years - From 'detecting' to 'preventing' eavesdropping" [27], which was named Round Robin DPS QKD protocol [28, 29].

Yuen comments furthermore on the quantitative security. The standard secret key length in QKD is about $|K| = 10^6$ bits, while the smallest $\varepsilon$ obtained experimentally [29] was

$$\Pr(\boldsymbol{K}\mid\boldsymbol{E})\sim\varepsilon=2^{-50}\gg\Pr(\boldsymbol{K})=2^{-|K|}=2^{-1{,}000{,}000}\tag{9}$$

This means the perfect secrecy $\Pr(\boldsymbol{K}|\boldsymbol{E}) = \Pr(\boldsymbol{K})$ cannot be satisfied for OTP, therefore the distributed key cannot guarantee the IID required for OTP encryption.

Surely, if $\varepsilon$ was small enough, $\Pr(\boldsymbol{K}|\boldsymbol{E})$ would correspondingly become small so Eve's success probability in eavesdropping would be small enough. However, $\Pr(\boldsymbol{K}|\boldsymbol{E}) \sim \varepsilon = 2^{-50}$ means its probability is as large as one car causes a fatal traffic accident in a year in Japan [19]. On the other hand, if there could not be any efficient algorithms to decrypt AES, the obtaining the correct key by Brute-force attack would be $2^{-256}$ for a 256-bit secret AES key. This would give easier understanding on a analogy repeatedly said by Yuen and Hirota that "QKD may be weaker than AES." With the current QKD systems, it is like operating $10^6$-bit OTP with an expanded 50-bit secret key.

It is often said that $\varepsilon$ could be arbitrarily small in theory, so we expect the future achievement by experiments, however, there is a trade-off that smaller $\varepsilon$ results in slower communication rate, and even in some cases there may be limitations in lowering $\varepsilon$ [30].

## 10.2 Known-Plaintext Attack on OTP after QKD

If IID of the distributed key cannot be satisfied, it is possible to gain the guessing probability of the rest of the part of the secret key by knowing some part of the key. Especially, OTP encrypts messages by XORing the plaintext and the key string, therefore if a part of the plaintext is possible to guess, the corresponding part in the key string is given immediately.

In conventional QKD theories, as it was described in Sec. 3, They can generate IID key with a probability of $1 - \varepsilon$, therefore there was no concern on Known-Plaintext Attacks (KPA). First of all, the reason why Ref. [3] introduced the trance distance was to prevent KPA by Eve with quantum memory, and Ref. [6] also said that KPA was proven to be impossible.

However, Yuen showed KPA is possible by the same derivation of (5) and estimated the probability [20]. In quantum way, it is proven as follows. Firstly, divide the secret key $\boldsymbol{k}_{\mathrm{A}}$ into a known part for Eve, $\boldsymbol{k}_{\mathrm{A1}}$, and the other unknown part, $\boldsymbol{k}_{\mathrm{A2}}$, then discard the quantum system corresponding to $\boldsymbol{k}_{\mathrm{A1}}$.

$$\rho_{\mathrm{ABE}}:=\sum_{\boldsymbol{k}_{\mathrm{A2}},\boldsymbol{k}_{\mathrm{B}}}\Pr(\boldsymbol{k}_{\mathrm{A2}},\boldsymbol{k}_{\mathrm{B}})\left|\boldsymbol{k}_{\mathrm{A2}},\boldsymbol{k}_{\mathrm{B}}\right\rangle\left\langle\boldsymbol{k}_{\mathrm{A2}},\boldsymbol{k}_{\mathrm{B}}\right|\otimes\rho_{\mathrm{E}}(\boldsymbol{k}_{\mathrm{A2}},\boldsymbol{k}_{\mathrm{B}})\tag{10}$$

$$\Gamma=\sum_{\boldsymbol{k}_{\mathrm{A2}}}\left|\boldsymbol{k}_{\mathrm{A2}},\boldsymbol{k}_{\mathrm{A2}}\right\rangle\left\langle\boldsymbol{k}_{\mathrm{A2}},\boldsymbol{k}_{\mathrm{A2}}\right|\otimes M_{\boldsymbol{k}_{\mathrm{A2}}}\tag{11}$$

In the same way in (8), calculate the guessing probability of the unknown part,

$$\Pr(\boldsymbol{K}_{\mathrm{A2}}\mid\boldsymbol{E})\le 2^{-|K_{\mathrm{A2}}|}+\varepsilon\tag{12}$$

Therefore, it is shown that KPA is not meaningless. However, the author personally thinks it does not have strong meaning compered to Ciphertext-Only Attack (COA), because $\varepsilon$ is far larger therefore the gaining of the first term in the right hand side in (13) would be negligible. In contrast, if $\varepsilon$ is small enough compared to $2^{-|K|}$, KPA will have stronger meaning.

## 10.3 Importance of Bit-Error-Rate for Eavesdropper

Yuen raised a question as follows: even if Eve could not obtain the correct key, but she obtained a key closed to the key Alice and Bob share, then what will happen? Eve cannot read the message at all if her key has just 1-bit error? How about 2 bits? Then 3 bits? Yuen emphasized the importance of Bit-Error-Rate (BER) for Eve, because it corresponds to the BER on the encrypted message, therefore he named it "BER Guarantee" [9, 20]. Clearly, a perfect key for OTP has IID so BER is always 1/2 for Eve, therefore she can never read the encrypted message. However, if she knows her BER is far smaller than 1/2, then she may be able to read some part of the encrypted message. Here is an example. Suppose you got a message "Tahnks" from your friend, you may think it was a typo of "Thanks," usually.

Now going back to the topic of QKD, here the author writes a rough estimation. Suppose Eve can read the message if her key has BER less than the certain BER, $Q_{\mathrm{E}}$. The number of such a situation is expressed as [31]

$$\sum_{a=0}^{|K|Q_{\mathrm{E}}}{}_{|K|}C_{a}\le 2^{|K|h_2(Q_{\mathrm{E}})}\tag{13}$$

Therefore, the rough estimation of Eve's success probability in obtaining a nearly correct message is

$$\Pr(K \mid E) 2^{|K|h_2(Q_E)} \leq 2^{-|K|(1-h_2(Q_E))} + \varepsilon 2^{|K|h_2(Q_E)} \quad (14)$$

Here, $h_2(Q_E)$ is a binary Shannon entropy with a probability of $Q_E$. As it is shown in (15), the chance Eve can read a nearly correct message would raise exponentially to the length of the secret key. So, if $\varepsilon = 2^{-50}$ and $|K| = 10^6$ bits, Eve's probability in obtaining a nearly correct message is almost $\Pr(\boldsymbol{K}|\boldsymbol{E}) = 1$ up to $Q_E = 2.5\times10^{-6}$, which means Eve's key has only a few bits of errors. Surely, even the author thinks the estimation by (14) is too rough, and Yuen himself wrote it is an open question that how much secure the message is under how much $Q_E$ is. We need further studies.

## 10.4 Criticisms on Derivation of Secure Key Rate

Yuen also questioned on the derivation of the secure key rate. The general procedures of QKDs are well known, but here the author describes as follows [31].

8. The transmitter Alice choose the bit to send and the encoding quantum basis randomly, then she sands a corresponding quantum state to the receiver, Bob.
9. Bob also chooses the measurement basis randomly, and obtain the classical bit from the measurement.
10. They repeat the above procedure, then they discuss on the classical public channel to discard the bits they chose different communication bases and holds the bits with the same communication bases.
11. Alice and Bob announces the part of their measurement results to estimate Quantum-Bit-Error-Rate, $Q$. If $Q$ is greater than the certain threshold, they abort the communication regarding they cannot yield secure key strings. When they can, they proceed to error-corrections in the key strings for key agreement.
12. Alice announces the parity check matrix for the error correction, and she calculates her syndrome with it. Then she sends her syndrome to Bob hiding it by OTP using the part of the pre-shared key.
13. Bob also calculates his syndrome using the parity check matrix Alice announced, then he operates error-correction comparing his syndrome with Alice's.
14. Finally, they proceed to Privacy Amplification to eliminate Eve's knowledge on the shared key, by announcing a hash function in public classical channel.

In the above process, the key consumption for OTP to hide Alice's syndrome is often given by

$$\mathrm{leak}_{\mathrm{EC}} = \xi k h_2(Q) \quad (15)$$

Here, $\xi$ is a factor chosen from 1 to 2, depending on the strength of the error correction code. Typically, it is set to $\xi = 1.1$ [31, 32].

To prove (15) for the case $\xi = 1$, see the following calculation. Condider $(k, m)$ linear codes, the total lengths are $k$ and the length of the information digits are $m$, which can correct up to $Qk$ errors. From the Hamming bound,

$$2^m \leq 2^k / \sum_{a=0}^{Qk} {}_kC_a \quad (16)$$

Then consider the following inequality.

$$\sum_{a=0}^{Qk} {}_kC_a \leq 2^{kh_2(Q)} \quad (17)$$

Therefore, the following inequality has to be satisfied.

$$\sum_{a=0}^{Qk} {}_kC_a \leq 2^{kh_2(Q)} \leq 2^{k-m} \quad (18)$$

Thus, the key consumption by OTP to hide Alice's syndrome is $k\, h_2(Q)$ bits.

However, Yuen explains as follows. If we use $(k, m)$ linear codes, the number of key candidates would shrink down to $2^m$ while we had $2^k$ possible candidates before the error correction. This problem would not be solved even if they hide the syndrome by OTP. To solve this problem, we have to consider $(n, k)$ linear codes by adding $n - k$ bits of parity check digits to the original key before error correction. Then, we have to consume $n\, h_2(Q)$ bits of the pre-shared key to hide the added parity check digit to tell Bob. The amount of $n$ is given by Hamming bound again,

$$\sum_{a=0}^{Qn} {}_nC_a \leq 2^{nh_2(Q)} \leq 2^{n-k} \quad (19)$$

Thus the key consumption by OTP for error correction is

$$\mathrm{leak}_{\mathrm{EC}} = kh_2(Q) / (1 - h_2(Q)) \geq kh_2(Q) \quad (20)$$

In addition, Yuen pointed out that choosing $\xi = 1.1$ habitually is not a "proven analysis" against QKD's original concept.

Here, the author gives some numerical analysis in Fig. 1 done in the study in [30]. When we use (20) derived by Yuen gives lower secure key rate especially in case of larger $Q$.

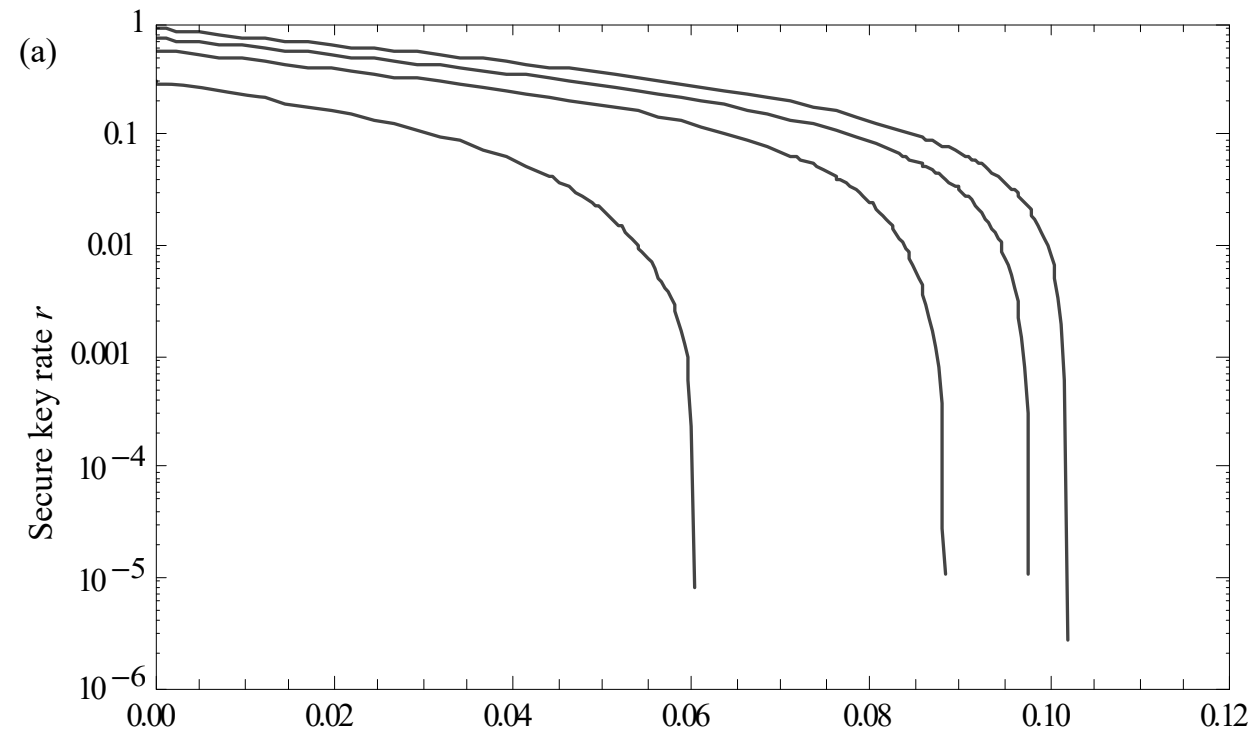

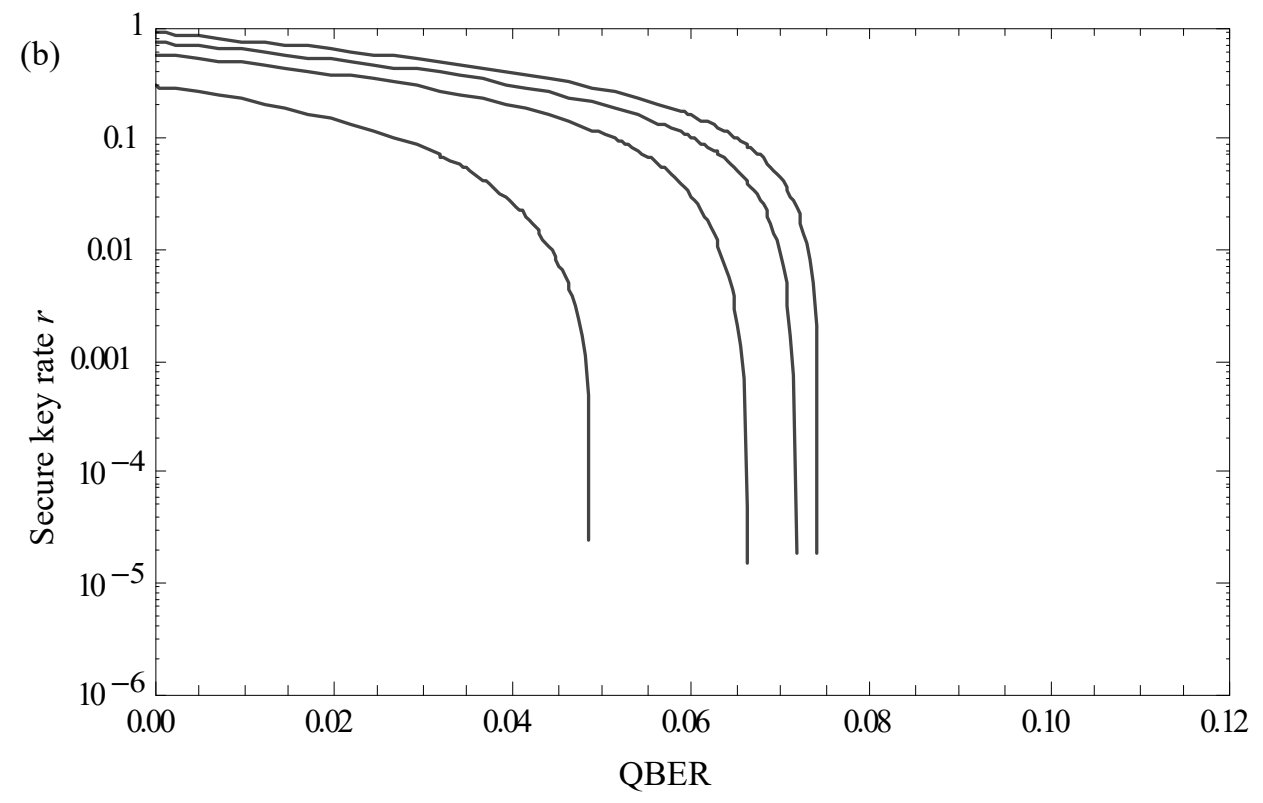

Fig. 1(a). Key rate with $\text{leak}_{\text{EC}}$ in (15), and (b) with $\text{leak}_{\text{EC}}$ in (20). From the lowest curve, the sifted key length = $10^5$, $10^6$, $10^7$, $10^9$ bits.

## 10.5 Criticisms on Effect of Privacy Amplification

Yuen also pointed out that Privacy Amplification may be rather harmful for the security of QKD. His description in [20] is not easy to understand, therefore, the author tries a different explanation. Consider Eve eavesdropped the quantum channel and store the states correlated to the legitimate users' key in her quantum memory. After Alice and Bob finished error correction, assume Eve measures her quantum memory and obtained the key string $\boldsymbol{k}_{\text{E}}$, while Alice and Bob share the key $\boldsymbol{k}_{\text{R}}$. Now, let Alice choose and announce a hash function from a set of a hash function family $F$, which is $\delta$-Almost Two-Universal as shown in (22), then

$$\Pr_{f \in F}(f(\boldsymbol{k}_{\text{E}}) = f(\boldsymbol{k}_{\text{R}})) \le \delta \tag{21}$$

There are two possible cases that Eve obtains the correct key $\boldsymbol{k}_{\text{E}} = \boldsymbol{k}_{\text{R}}$, and $\boldsymbol{k}_{\text{E}} \neq \boldsymbol{k}_{\text{R}}$ but collision occurs because of the property of hash functions. Therefore, Eve's success probability in obtaining the correct key in the end is,

$$\begin{aligned} P_{\text{guess}} &= \Pr(\boldsymbol{k}_{\text{E}} = \boldsymbol{k}_{\text{R}}) + \Pr(\boldsymbol{k}_{\text{E}} \neq \boldsymbol{k}_{\text{R}}) \Pr(f(\boldsymbol{k}_{\text{E}}) = f(\boldsymbol{k}_{\text{R}})) \\ &\le \Pr(\boldsymbol{k}_{\text{E}} = \boldsymbol{k}_{\text{R}}) + \left(1 - \Pr(\boldsymbol{k}_{\text{E}} = \boldsymbol{k}_{\text{R}})\right)\delta \\ &= (1-\delta)\Pr(\boldsymbol{k}_{\text{E}} = \boldsymbol{k}_{\text{R}}) + \delta \end{aligned} \tag{22}$$

The last part of (22) is larger than Eve's guessing probability before the Privacy Amplification. Therefore the following question arises: does Privacy Amplification really gain the security of the distributed key? Consider the following example. Let $|K_{\text{R}}|$ be the key length before Privacy Amplification, let $|K|$ be the key length after Privacy Amplification, therefore $|K_{\text{R}}| > |K|$. If Eve does not even eavesdropping on the quantum channel, but she guesses the correct key by pure guessing. It is trivial that she has larger probability of guessing the correct key after hashing than she has before hashing. Yuen explained that the reason why Privacy Amplification has given misconception that it would enhance the key security was, that the averaging the hashing performance over the hashing family in Leftover Hash Lemma. However, in reality, Alice announces publically which hash function they use, therefore Eve exactly knows which will be used. Therefore, to evaluate the performance of Privacy Amplification, we need to evaluate the performance of a chosen hash function without averaging.

A more complicating problem is related to the previous topic of error correction. If we regard the sifted key as $(k, m)$ linear codes, then there should be correlations among key bits, because we regard $k$-bit key as $(k, m)$ code, there are only $2^m$ patterns of key candidates instead of $2^k$ patterns of key candidates. Evaluating the effect of Privacy Amplification is not easy when there are correlations between key bits. Therefore, again, we need to add parity check digits to the sifted key to make it $(n, k)$ code, to have less correlations among key bits, so we have to take the previous problem seriously.

## 10.6 Authenticity of Communication Channels

There seem to be some people misunderstanding outside of the QKD researcher community because it is a common sense among QKD researchers, thus it is rarely explained, so the author explicitly writes here. Before starting QKD, Alice and Bob need to have pre-shared authentication key to recognize each other [31, 33, 34]. Otherwise, Eve can launch Man-in-the-Middle Attacks by she pretends to be Bob to Alice, and same for Bob to be Alice, relaying both classical and quantum signals coming from Alice to Bob, which allows not only total eavesdropping but also falsifying the messages. In this sense, QKD is not a public key distribution technology to replace public key encryptions like RSA, say, the public key of RSA is known to even Eve, but the authentication key in QKD should not be disclosed to Eve. Therefore, Alice and Bob need to share the authentication key secretly in some way. In this meaning, QKD is similar to symmetric key cryptography like AES, unlike public key encryptions like RSA.

Yuen pointed out the security level of this authentication key. We may be able to share an authentication key with IID, but what will happen if we renew the authentication key by the part of the distributed key? As it is explained in Sec. 4, Eve guesses the correct key with a probability of about $\varepsilon$. Yuen regards the security of authentication is far more important than the security level of encryption, therefore he claims $\varepsilon$ has to be much smaller than we currently can obtain. Even if $\varepsilon$ is small enough, the renewed authentication key is a part of the distributed key which is known by Eve with a probability of about $\varepsilon$, resulting in security degradation compared to the initial authentication key with IID, and this continues as far as QKD operation is continued. Furthermore, the part of the distributed key which is known by Eve with a probability of about $\varepsilon$ has to be used in OTP to hide the syndrome explained in Sec. 4.4. Therefore, the influence of the degradation of the security level of the distributed key has to be evaluated to claim the provable security.

## 11 Recent trends of Quantum Cryptography

Defense Advanced Research Projects Agency (DARPA) in USA announced the requirement to quantum cryptography in 2012 as follows [35].

・Communication speed: 1-10 Gbps
・Communication range: 1,000-10,000 km

These are seriously challenging goals for QKD.

Meanwhile, Communications-Electronics Security Group (CESG) in UK distributed a white paper to suggest withdrawal of QKD developments [36]. The reasons are as follows.

6. QKD is mainly targeting message encryptions, but the applications for signature, authentication or any other use in security have barely been studied.
7. QKD is basically Point-To-Point communication, and it has limitations in communication range. Furthermore, to construct the network, it must have network nodes not protected by quantum mechanics, or other provable ways, therefore it cannot promise provable security.
8. There are some studies of attacks on QKD systems [23, etc.], however, these studies will not reveal all potential threat on the systems. There would always be unknown loopholes remained.
9. There are alternatives to QKD. For example, quantum-resistant cryptographies are implemented by software, while QKD needs to be implemented by hardware, resulting in hardware replacements to update.
10. In conclusion, it is recommendable to develop quantum-resistant cryptographies implemented by software.

On the other hand, there are other quantum cryptographies than QKD. Firstly, Yuen himself proposed a protocol named Keyed Communication in Quantum-noise (KCQ) [21] (which is called Y-00 [22] or quantum noise stream cipher, especially customized one in Tamagawa Univ. is named "Quantum Enigma Cipher" [37]), S. Lloyd proposed Quantum Enigma Machine [38], J. H. Shapiro proposed a quantum cryptography protocol using quantum illumination technology [39]. These protocols do not use weak signals like single photons, but use macroscopic quantum nature with intensity of current optical communication, therefore they would satisfy DARPA's requirements.

Now, limiting the topic on KCQ, it is often misunderstood that it is a technology to encrypt messages directly using the initial key, not a technology to distribute secret keys. However, it is possible like QKD to distribute secret keys by replacing the message to be sent by the secret key to be shared. Furthermore, it has often been proposed to combine QKD and KCQ to distribute the secret key for KCQ to communicate on it. However, there may not be much meaning for KCQ if the security of QKD is strengthen furthermore.

It is sure that the QKD researchers those who have pioneered highly secure communication using quantum mechanics have to be greatly respected. Adding to it, the problems appeared in QKD researches have shown important challenges that what is required for physical encryption systems. However, from the view point of practical use, there would be choice to study other quantum cryptographies other than QKD. Even in such a case, knowledge obtained in QKD studies will be greatly useful, in the author's opinion.

## 12 Summary

Quantum Key Distribution (QKD) has been attracting many researchers because it would make possible to distribute secret keys in provable ways since its invention in 1984. However, in 2009, H. P. Yuen have arisen questions on its security claims, then O. Hirota, K. Kato, T. Iwakoshi have flowed Yuen's studies, and some explanations why Yuen's claims were correct. Then, Yuen himself published a paper in 2016 summarizing his criticisms he proposed, pointing out what the gaps are between security claims of QKD and actual situation. This article explained problems Yuen has pointed out in details, then recent trends in quantum cryptographies exploring other protocols than QKDs. Indeed, QKD itself has an important meaning that it was the first step that showed highly secure communication is possible using quantum mechanics. However, it has been clarified by Yuen that QKD still has problems to claim provable security, and choices have been proposed other than QKD. Even in case, the knowledge obtained in studies of QKD would be useful, in the author's opinion. And if researchers would still continue studies on QKD, the author wishes that they would take Yuen's criticisms very seriously.